\documentstyle[sprocl,epsfig]{article}

\newcommand{\be}{\begin{equation}}
\newcommand{\ee}{\end{equation}}
\newcommand{\ba}{\begin{array}}
\newcommand{\ea}{\end{array}}

\def\be{\begin{equation}}
\def\ee{\end{equation}}
\def\bea{\begin{eqnarray}}
\def\eea{\end{eqnarray}}

\begin{document}


\title{RECONSTRUCTING NEUTRINO MASS SPECTRUM
\footnote{Talk given at  
the 5th International WEIN Symposium: 
A Conference on Physics Beyond the Standard Model (WEIN 98),
Santa Fe, NM, June 14--21, 1998.}
}

\author{A. Yu. SMIRNOV}

\address{The Abdus Salam International Center of Theoretical Physics,\\ 
Strada Costiera 11, Trieste, Italy \\
Institute for Nuclear Research, RAS, Moscow, Russia \\
E-mail:  smirnov@ictp.trieste.it}


\maketitle\abstracts{Reconstruction of the neutrino mass 
spectrum and lepton mixing is one of  the fundamental 
problems of particle physics. In this connection we consider 
two central topics: (i) the origin of large 
lepton mixing, (ii) possible existence of new (sterile) 
neutrino states. 
We  discuss also possible relation between large mixing and 
existence of sterile neutrinos.}

\section{Introduction }

The experimental situation  can be
summarized in the following way:

1. Recent SuperKamiokande (SK)  results 
on the atmospheric neutrinos  give strong evidence for the 
oscillations of the muon neutrinos with large (maximal) depth \cite{SKat}. 
An   open question is to which extend  the electron neutrinos 
are involved in the oscillations 
and whether an excess of the e-like events exists. 

2. The situation with solar neutrinos is rather uncertain. 
The data indicate unexpected  distortion of the 
 recoil electrons energy spectrum \cite{SKsun}.   
It is unclear whether we deal with just statistical fluctuations, or   
{\it distortion} of the boron neutrino spectrum or  
an  {\it excess} of the events near the end point 
which is not related to boron neutrinos. 
No day-night asymmetry and no 
earth core enhancement of signal have been found.

3. LSND collaboration has  further confirmed the 
oscillation interpretation of their result \cite{LSND}. 
At the same time,  KARMEN \cite{KAR} does not see  the  oscillation effect
concluding that the data  are approaching the  situation 
when one can speak on direct
contradiction between the two experiments.  

4. Recent cosmological observations 
(early galaxies, clusters evolution, high redshift supernova type IA 
data) show that a contribution of neutrinos to the energy 
density of the Universe should be smaller than it was thought earlier, and 
the Hot Dark Matter (HDM)  is 
not necessary for the  fit of  data on the large scale structure
\cite{HDM}. 
At the same time, some amount of the HDM 
is not excluded and may be needed for the further tuning of the 
data. \\

Keeping this in mind, 
we will concentrate on  models which 
explain the solar and the atmospheric neutrino data. 
We will consider main issues 
of the present day discussions: 

1. Origin of the large leptonic mixing.  

2. Possible existence of new neutrino states (sterile 
neutrinos). 

3. We also comment on possible relation of these two issues,  
addressing the question:
is large mixing  the  mixing with sterile neutrinos?\\ 
 
According to the SK  result,   muon neutrinos oscillate into 
tau neutrinos or probably into sterile neutrinos. The effective 
mixing angle which determines the depth of oscillations 
should be large in both cases
\be 
\sin^2 2\theta > 0.8. 
\ee
The favoured mode is $\nu_{\mu} - \nu_{\tau}$ \cite{}, although 
$\nu_{\mu} - \nu_s$ \cite{} gives comparably good fit of the data. 
Pure $\nu_{\mu} - \nu_e$ channel is strongly disfavored by the
SuperKamiokande data itself \cite{},  
and restricted by the CHOOZ result~\cite{CHOOZ}. At the same time, 
a small contribution of the $\nu_{\mu} - \nu_e$ channel 
is possible and probably desired in view of some excess of the 
$e$ - like events.\\ 

In this connection the basic 
questions are 

\begin{itemize}

\item
{\it Why lepton mixing is large while 
quark mixing is small?} Is this consistent with 
quark-lepton symmetry (correspondence) and Grand Unification?  
The question has more general  conceptual nature.  
The picture we had before is that known quarks and leptons   
form families with weak interfamily connection 
(characterized by mixing). Should we support 
this conception in view of
maximal mixing between the second and the third 
generations of leptons? 

\item
{\it Is lepton mixing maximal between the 
second and the third generations only, or probably all lepton mixings 
are large?} In other words is the  observed large lepton mixing the
feature of the second and third generation
or it  is the property of all leptons?   

The answer to this question will come from studies 
of solar neutrinos.  In the first case  the  small mixing 
MSW - solution is realized, whereas in the second case 
the choice will be between the large mixing MSW solution and  
long range vacuum oscillations (``just-so").  

\end{itemize}
 
Completely different possibility is that large mixing is the 
mixing with new (sterile) neutrino state. 
In this case the mixing between
flavor states can be  small in analogy with quark 
mixing. 

\section{Patterns of neutrino mass and mixing}

Before going into details of the theoretical analysis, we will describe
possible patterns of the neutrino mass and mixing 
which are implied by phenomenology. Here we consider three types of 
neutrino schemes with single, double (bi-) and 
triple maximal mixing.

\subsection{Single maximal mixing}
 
\begin{figure}[htb] \hbox to
\hsize{\hfil\epsfxsize=7cm\epsfbox{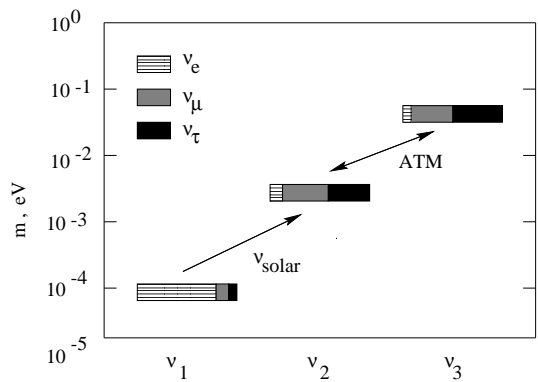}\hfil} 
\caption{~~Neutrino masses 
and mixing  for the ``solar and atmospheric"  neutrinos.
Boxes correspond to the mass eigenstates. The sizes of different
regions in the boxes show admixtures of different flavors. Weakly hatched
regions correspond to the electron flavor, strongly hatched regions depict
the muon flavor, black regions present the tau flavor.  } 
\label{solat}
\end{figure}

The scheme has the hierarchical mass spectrum 
\be 
m_3 = (0.3 - 3) \cdot
10^{-1} {\rm eV},~~~ m_2 = (2 - 4) \cdot 10^{-3}{\rm eV},~~~ m_1 \ll m_2
\label{ma}
\ee 
with $\nu_{\mu}$ and $\nu_{\tau}$ mixed strongly in $\nu_{2}$ and
$\nu_{3}$ (see fig. \ref{solat}). The electron flavor is weakly mixed:  
it is mainly
in $\nu_{1}$ with small admixtures in the heavy states. 
The solar neutrino data are explained by  
$\nu_e \rightarrow \nu_{2}$ resonance conversion inside the Sun. 
Notice that $\nu_e$ converts
to $\nu_{\mu}$ and $\nu_{\tau}$ in comparable portions.  
The atmospheric neutrino problem  is solved via
$\nu_{\mu} \leftrightarrow \nu_{\tau}$ oscillations. Small $\nu_e$
admixture in $\nu_{3}$ can lead to resonantly
enhanced oscillations in  matter of
the Earth.

There is no explanation of  the LSND result,  and
the contribution to the Hot Dark Matter component of the universe is
small:  $\Omega_{\nu} < 0.01$.  

The scheme can provide significant amount of the HDM  
without change of the oscillation 
pattern if all three neutrinos have degenerate masses: 
$m_i \approx m_0 \sim 1$ eV  with small splitting:   
\be
\Delta m_{12} \approx \frac{\Delta m^2_2}{2 m_0} = 
\frac{\Delta m^2_{\odot}}{2 m_0},  ~~~~~ 
\Delta m_{23}\approx  \frac{\Delta m^2_3}{2 m_0} = 
\frac{\Delta m^2_{atm}}{2 m_0},   
\ee 
where $m_2$ and $m_3$ are defined in (\ref{ma}) and   
\be 
\Delta m^2_{\odot} \approx 6 \cdot 10^{-6} {\rm eV}^2,  ~~~~~~
\Delta m^2_{atm} \approx (10^{- 3} -  10^{- 2}) {\rm eV}^2 . 
\ee
In this case an  effective Majorana mass of the electron 
neutrino equals $m_{ee} \approx m_0$ and searches for the 
neutrinoless double beta decay give crucial check of 
the scheme.

The scheme can be probed by  the long baseline experiments.

\subsection{Bi-large and bi-maximal mixings}

The previous scheme (fig.~\ref{solat}) can be modified in such a way that 
solar neutrino data  are explained by large angle MSW conversion 
with $\sin^2 2\theta \sim 0.7 - 0.9$ and 
$\Delta m^2 = (2 - 20) ~ 10^{-5}$ eV$^2$.   

In a version of the scheme with mass degeneracy, 
the cancellation in the
effective Majorana mass of the electron  
neutrino can occur,  so that $m_{ee} \sim m_0 \sqrt{1 - \sin^2 2\theta}$ 
will  be substantially lower than present bound 
even for $m_0 > 1$ eV (see \cite{MY} for recent discussion).

The solution of the atmospheric neutrino problem is basically 
the same as in the previous case. 
There is a suggestion \cite{mkkee} that mass splitting 
$\Delta m^2 >  10^{-4}$ eV$^2$ between the two light states 
could be relevant for the atmospheric neutrino problem. 
In particular, this mode can lead to 
the zenith angle dependence of the detected  events.  
In this case the 23-splitting could be much larger to accommodate 
the LSND result. 
However, in \cite{mkkee}  the matter effect has not been taken into
account, and  the latter,    
in  fact, strongly suppresses the oscillation depth.\\ 
\begin{figure}[htb]
\hbox to \hsize{\hfil\epsfxsize=7cm\epsfbox{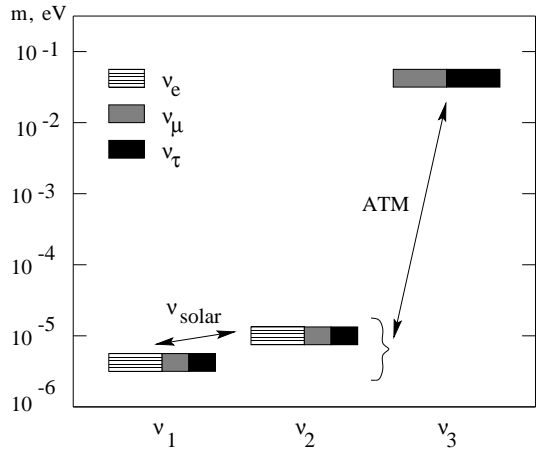}\hfil}
\caption{~~Neutrino masses and mixing pattern  of the  bi-maximal 
mixing  scheme.
}
\label{fbimax}
\end{figure}

In the bi-maximal scheme  \cite{bimax} neutrinos have masses 
\be
m_3 = (0.3 - 3) \cdot 10^{-1} {\rm eV},~~~
m_2 \sim (0.7 - 2) \cdot 10^{-5}{\rm eV},~~~
m_1 \ll  m_2
\ee
(see fig. \ref{fbimax});
$\nu_{\mu}$ and $\nu_{\tau}$ mix maximally in   
$\nu_{3} = (\nu_{\mu} + \nu_{\tau})/\sqrt{2}$; the orthogonal
combination, $\nu_2' \equiv (\nu_{\mu} - \nu_{\tau})/\sqrt{2}$
strongly mixes with
$\nu_e$ in $\nu_{1}$  and  $\nu_{2}$.
There is no admixture of $\nu_e$ in the $\nu_{3}$.
The corresponding mixing matrix has the form:  
\be
V_{MNS} =
\left(
\matrix{
\frac{1}{\sqrt{2}}  & - \frac{1}{\sqrt{2}} & 0                   \cr
\frac{1}{2}  & \frac{1}{2} & - \frac{1}{\sqrt{2}}  \cr
\frac{1}{2}  & \frac{1}{2} &  \frac{1}{\sqrt{2}}   \cr}
\right),
\label{viss}
\ee
\begin{figure}[htb]
\hbox to \hsize{\hfil\epsfxsize=7cm\epsfbox{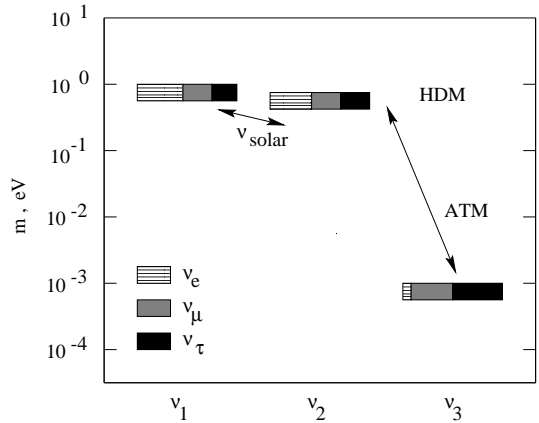}\hfil}
\caption{~~The neutrino mass and mixing  in  the  bi-maximal
mixing  scheme with inverse  mass hierarchy.
}
\label{fbimaxinv}
\end{figure}

The solar neutrino problem  can be solved via
$\nu_e \leftrightarrow \nu_2'$  ``Just-so" vacuum oscillations.
Notice
that $\nu_e$ converts equally to $\nu_{\mu}$ and $\nu_{\tau}$. 
Larger values of $\Delta m^2$ lead to the averaged oscillation result 
which does not give a good fit of the solar  neutrino data.  

The atmospheric neutrino anomaly is solved via
$\nu_{\mu} \leftrightarrow \nu_{\tau}$ maximal depth oscillations.

Let us comment on the version of the
bi-maximal scheme with inverted mass
hierarchy:  
$
m_1 \approx m_2 \gg m_3 ~  
$
(\ref{fbimaxinv}). 
Such a possibility can be realized in the model
with approximate $L_e - L_{\mu} - L_{\tau}$-symmetry.  
The corresponding  mass matrix  has the form: 
\be
m_{\nu} =
\left(
\matrix{
\epsilon' & 1 & 1                  \cr
1  & \epsilon & \epsilon  \cr
1  & \epsilon  & \epsilon    \cr}
\right). 
\label{lmutau}
\ee
Two states with maximal (or large) $\nu_e$ mixing
are heavy and degenerate, whereas the third state  with large
$\nu_{\mu} - \nu_{\tau}$ mixing and small $\nu_e$ admixture  is light.
In this scheme the $\nu_e - \nu_3'$ level crossing
occurs in the {\it antineutrino} channel, so that
in supernovae  $\bar{\nu}_e$ will be strongly converted into
combination of $\bar{\nu}_{\mu}$, $\bar{\nu}_{\tau}$ and vice versa.
As the result the $\bar{\nu}_e$'s will have hard spectrum 
of the original $\bar{\nu}_{\mu}$.

One can introduce a  degeneracy of neutrinos
(keeping the same $\Delta m^2$) to get significant amount the HDM
in the Universe without change of the oscillation pattern. The
effective 
Majorana mass of the electron neutrino is zero 
in the strict bi-maximal case, so that  no effect 
in the double beta decay is expected due to light  neutrinos.   
In this scheme  one needs two  mass splittings of the order 
$10^{-3}$ eV and 
$10^{-10}$ eV respectively which looks very unnatural.

\subsection{Threefold maximal mixing}

\begin{figure}[htb]
\hbox to \hsize{\hfil\epsfxsize=7cm\epsfbox{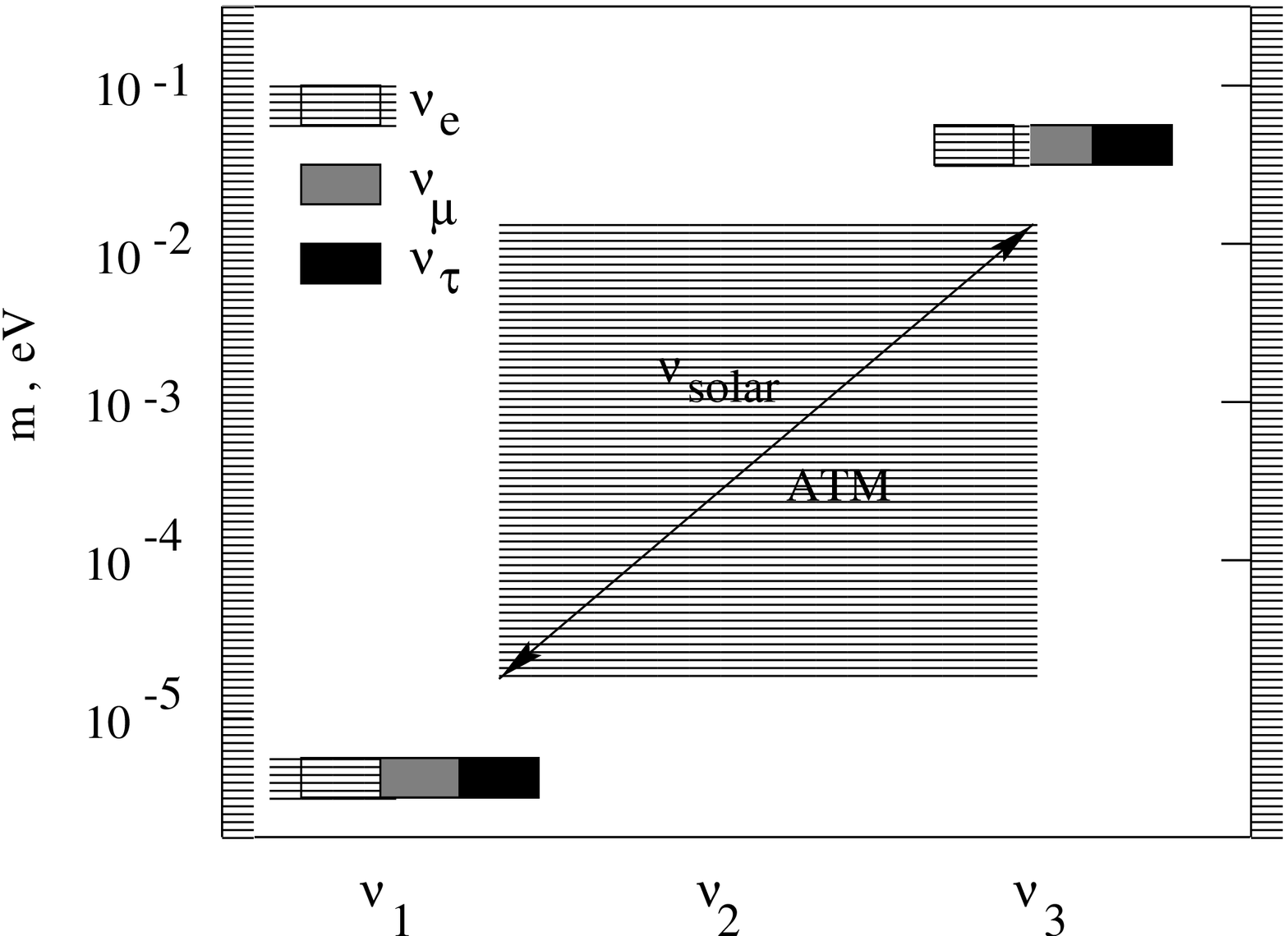}\hfil}
\caption{~~Neutrino masses and mixing  in  the  
scheme with threefold maximal mixing.
}
\label{ftriple}
\end{figure}

In such a scheme \cite{three} all the elements of the mixing matrix are
assumed to be equal: $|U_{\alpha i}| = 1/\sqrt{3}$ (\ref{ftriple}). 
In all flavor channels 
All three frequencies of oscillations contribute
to  all flavor channels equally.  
The atmospheric neutrino problem is solved by 
$\nu_{\mu} \leftrightarrow \nu_{e}$ and 
$\nu_{\mu} \leftrightarrow \nu_{\tau}$ oscillations 
with equal depth: $\sin^2 2\theta = 4/9$, so that the $\nu_{\mu}$ -
disappearance is characterized by $\sin^2 2\theta = 8/9$. 
The CHOOZ bound implies that $\Delta m^2 < 10^{-3}$ eV$^2$. 

The solar neutrino survival probability 
equals $P = 4/9 P_2 + 1/9$, where $P_2$ is the two 
neutrino oscillation probability with maximal depth and 
smallest mass splitting $\Delta m_{12}^2$. 
It is assumed that     $\Delta m_{12}^2 < 10^{-11}$ eV$^2$,  
so that  1 - 2 subsystem of neutrinos is frozen and $P_2 = 1$. 
As the result the solar neutrino flux will have 
energy independent suppression $P = 5/9$. 
The fit of both the  atmospheric and the solar neutrino data 
is substantially worser than in previous schemes.

\section{Large Lepton Mixing}
\label{LLM}

\subsection{Is large lepton mixing the  problem?}

Let us first clarify  whether the problem of large mixing 
exists at all.  The conception of families of  fermions 
can be expressed in the following way.   
In certain basis  mass matrices of both upper  and
down  fermions (from  doublets)  
have hierarchical structure with small
off-diagonal elements. The matrices are considered  to be  natural
\cite{Peccei} if the mixing angles 
$\theta_{ij}$ satisfy inequality 
\be 
|\theta_{ij}| \leq \sqrt{\frac{m_i}{m_j}}~, 
\label{natangle}
\ee
where $m_i$ and $m_j$ are the eigenvalues. 
(In this case no special arrangement of  the matrix 
elements is needed).  

Let us consider the second and third  generations of  
leptons and  introduce 
the angles $\theta_{cl}$ and  $\theta_{\nu}$ 
which diagonalize the mass matrices of the charged leptons and the 
neutrinos correspondingly. 
Then the lepton mixing angle equals 
\be
\theta_l = \theta_{cl} - \theta_{\nu}~. 
\label{lmix}
\ee
Using (\ref{natangle}) we get
\be
\theta_{cl} \approx  \sqrt{\frac{m_{\mu}}{m_{\tau}}} \sim 13^{0},~~~  
\theta_{\nu} \approx   \sqrt{\frac{m_2}{m_3}} \sim 14^{0}~, 
\label{lmix}
\ee
where for $m_3 \sim 0.05$ eV and $m_2 \sim 0.003$ eV are the  values 
of masses required by  solutions of the atmospheric 
and the solar neutrino problems. 
If the angles   $\theta_{cl}$ and  $\theta_{\nu}$ have 
opposite signs, so that $\theta_l = |\theta_{cl}| + |\theta_{\nu}|$, 
we find $\theta_l = 27^{\circ}$ and $\sin^2 2\theta = 0.67 - 0.72$ - 
close to the desired value~\cite{barshay}. 
Thus, the  large lepton mixing is consistent with 
the naturalness of the mass matrices. 

Notice that if neutrino masses are due to the see-saw mechanism 
and the mass matrix of the RH  neutrinos 
has no hierarchy: $ \sim M_0 \cdot \hat{I}$, then 
$m_i \propto m_D^2$ and the mixing angle is determined by the 
Dirac mass matrix $m_D$. In this case 
relation between the masses and  mixing becomes 
$|\theta_{\nu}| \approx~ ^4\sqrt{{m_2}/{m_3}}$~  \cite{FY} 
which leads to  $\sin^2 2\theta = 0.96$. 

For quarks the mixing is small if  
the corresponding angles $\theta_{u}$ and $\theta_{d}$ 
have the same signs and therefore cancel each other in the total mixing.  
The same cancellation may occur for the mixing of the first and second
generations, thus leading to a small mixing solutions of the solar
neutrino problem. 

So,  the problem of the large mixing is reduced to explanation 
of signs (phases in general) 
of contributions to mixing from the upper and the down fermions. 
In fact, the change of the relative sign of the contributions 
in the lepton sector can be related to the see-saw origin of the neutrino
mass \cite{tao}. 

Thus, the  large lepton mixing 
can be well consistent with our ``standard notions": 
quark -  lepton symmetry (similarity of the Dirac mass matrices), 
usual family structure and the see-saw mechanism. 

The alternative possibility is that  
large lepton mixing is a manifestation of  new physics beyond 
the ``standard notion".  In what follows we will concentrate on this
interpretation. 

\subsection{Classifying possibilities}

Trying to answer the question why the lepton mixing is large, 
while the quark mixing is small one can think about the following 
possibilities: 

\begin{itemize}

\item 
Large lepton mixing is the mixing of muon neutrino with 
sterile neutrino. In this case the question does not exist: 
There is no analogue of  $\nu_{\mu} - \nu_s$ mixing in the quark 
sector. 

\item
If  the atmospheric neutrino anomaly is due to 
$\nu_{\mu} - \nu_{\tau}$ mixing,  there are two options: 

\end{itemize}

1). Mechanism of the neutrino mass generation  
differs from that of the quarks. For instance, the mass matrix could be  
\be 
m_{\nu} = m_{\nu}^{rad} + m_{\nu}^{see-saw}~, 
\ee
where $m_{\nu}^{see-saw}$ is the see-saw 
contribution,  whereas  $m_{\nu}^{rad}$ is the contribution from 
radiative  mechanism. 
The radiative contribution can dominate and the 
role of the see-saw is just to suppress the 
effect of the Dirac mass term. 
The simplest version of the radiative mechanism which leads to a  
large lepton mixing is the Zee-mechanism \cite{zee}. 
The key element is new charged scalar  boson $S^+$ being singlet of 
$SU(2)$. (Also second higgs doublet is introduced to have  the 
couplings with  $S^+$). 
In this case large lepton mixing is the consequence of 
 
-~~ SU(2) gauge symmetry: the coupling 
of the singlet with lepton doublets 
$f_{\alpha \beta} L_{\alpha}^T i\sigma_2 L_{\beta} S^+$
is  antisymmetric in family index. 

-~~ assumption that there is no strong inverse hierarchy of 
  $f_{\alpha \beta}$, 

-~~ mass hierarchy of charge leptons. 

The model can be supplied by additional sterile neutrino to explain the 
solar neutrino problem \cite{TS}.\\

2). Mechanism of the neutrino mass and lepton mixing 
generation is closely related to  generation 
of the quark and charged lepton masses. 
This possibility is realized by the see-saw mechanism \cite{seesaw}. 
According to the see-saw mechanism: 
\be
m_{\nu} = - m_{\nu}^D M^{-1} m_{\nu}^{D~ T} + m_0, 
\ee 
where $m_{\nu}^D$ is the Dirac mass matrix of neutrinos, 
$M$ is the Majorana matrix of the RH components and 
$m_0$ is the direct majorana mass matrix of the left components 
which appears if the scalar ($SU_2$) triplet exists with  
no-zero  VEV. 

In  the quark sector  the mixing is determined 
by two matrices: $m_u$ for the upper quarks and 
$m_d$ for the down quarks. The mixing (CKM -) matrix  
is the product of matrices of the left  component rotations:  
\be 
V_{CKM} = V_u^{\dagger} \cdot V_d . 
\ee
In contrast, the lepton mixing is determined by three matrices 
$m_{\nu}^D$, $m_{l}$ and $M$, and the lepton (MNS) 
mixing matrix \cite{MNS} can be written as: 
\be
V_{MNS} = V_{ss} \cdot V_{\nu} \cdot V_l~.  
\label{lmix}
\ee
Here $V_{ss}$ is the see-saw matrix which specifies the 
see-saw mechanism itself \cite{AS}. It describes the influence of the 
matrix $M$ structure on the lepton mixing. 
Obviously, if $M \propto I$,  $V_{ss} = I$  and 
$V_{MNS} = V_{\nu}^{\dagger} \cdot V_l$  in   analogy with  the CKM
structure. 
According to  (\ref{lmix}), there are three possible 
sources of the large mixing (of course, the interplay of several is
possible):  

\begin{itemize}

\item $V_{ss}$, that is,  the see-saw mechanism itself 
leads to enhancement (the see-saw enhancement);  

\item $V_{\nu}$ which follows  from Dirac neutrino mass matrix;  

\item $V_l$ which follows from mass matrix of the charged leptons.  

\end{itemize}

Here we have neglected possible effect of $m_0$, which in fact 
can also be important. 
Notice, that precise origin of the enhancement 
({\it e.g.}   $V_{\nu}$ or  $V_l$)  depends on basis in which
the mass matrices are introduced. 

Recently, a number of models have been suggested 
which realize the three above  possibilities.

\subsection{See-saw enhancement of lepton mixing}

Not only the smallness of the neutrino mass 
but also large lepton mixing  can be related to  Majorana 
nature of neutrinos and both can follow from the see-saw mechanism. 

It is natural to assume (in a spirit of the grand unification) 
that the lepton 
Dirac  mass matrices are similar to the quark mass matrices 
at some unification scale: 
$m_{\nu}^D \sim  m_u$ and $m_l \sim m_d$, and moreover,  for the third 
generation one may expect the  equalities: 
$m_{\nu 3} = m_t$, $m_{\tau} = m_b$. 
Then the difference in the quark and lepton mixing  
can follow from specific structure of $M$ \cite{}. 

If the influence of the first generation on the 
mixing of the second and the third generations  is small  
(and the problem is reduced to  two generation problem),  
one gets two different conditions of the strong 
see-saw enhancement \cite{sees}: 

(i) Strong interfamily connection.   
In the basis where the neutrino Dirac mass matrix is diagonal 
(Dirac basis), $M$ should be off-diagonal: 
\be
M \sim  M_0 \left(\matrix{ a & 0 & 0 \cr
                       0 & 0 & b \cr
                       0 & b & 0 \cr} \right)~,  
\label{mat1}
\ee
where $a$ and $b$ are some numbers.  
The off-diagonal form of $M$ can in turn  be related to the  
Majorana nature of neutrinos. Prescribing the horizontal 
charges (0, 1, -1) we reproduce  (\ref{mat1}). 

\begin{figure}[htb]
\hbox to \hsize{\hfil\epsfxsize=6cm\epsfbox{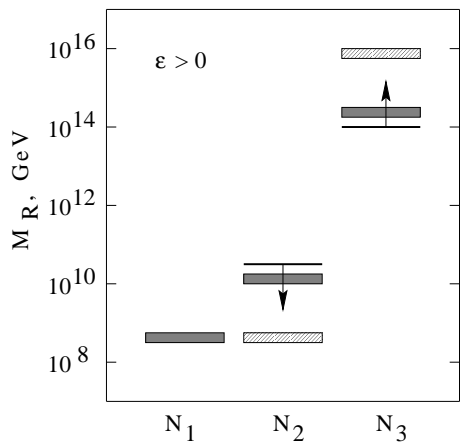}\hfil}
\caption{Masses of the RH neutrinos 
for the solar and atmospheric neutrinos. Solid lines correspond to 
masses in absence of mixing in the Dirac basis, $M_{0i}$. 
Shadoved boxes correspond to small (natural value of) mixing.  
Product of masses $M_2 \cdot M_3$ does not depend on  
mixing.  Large mass hierarchy (weakly shadowed boxes)  
leads to enhancement of mixing.}  
\label{fscale}
\end{figure}

(ii) Strong mass hierarchy. 
In the two generation case the 
matrix $V_{ss}$ can be parametrized by  
the see-saw angle $\theta_{ss}$ which can be 
related to  the hierarchies of the eigenvalues  
of the Dirac, $m_i^D$, and Majorana, $M_i$,  mass matrices: 
$\epsilon_D \equiv m_2^D/m_3^D$,   
$\epsilon_M \equiv M_2/M_3$.  
Introducing also 
$\epsilon_0 \equiv M_{02}/M_{03} \equiv  \epsilon_D^2 m_3/m_2$ 
(where $M_{0i}$ are the masses of the RH neutrinos which give 
in the absence of mixing in $M$ the masses of the light neutrinos 
$m_2$ and $m_3$) we get~\cite{sees} 
\be
\sin \theta_{ss} \approx \frac{\epsilon_D^2}{\epsilon_0}
\left(\sqrt{\frac{\epsilon_0}{\epsilon_M}} -1 \right)~.  
\label{ssangle}
\ee
Clearly, without mixing in $M$ in the Dirac basis 
$\epsilon_M = \epsilon_0$ and  
the $\theta_{ss} = 0$. We get  
$\theta_{ss} \sim 1$, if 
$\epsilon_M \sim \epsilon_D^2 \sim 10^{-6}$,  so that 
one mass can be  $M_2 \sim 10^{9}$ GeV and another one     
$M_3 \sim 10^{15}$ GeV (see fig. \ref{fscale}). 
This opens interesting possibility, that 
third neutrino acquires the $GU$ -scale mass. 
Two other RH neutrinos are  massless at $\Lambda_{GU}$  
and acquire  masses at the intermediate scale. 
Notice that in this case,  mixing in the $M$ should
be relatively large: 
\be
\sin^2 \theta_M = \sqrt{\epsilon_0 \epsilon_M}~. 
\ee

Recently such a possibility has been realized in the 
scenario with 
textures of the mass matrices which also describe the masses of the
charged fermions \cite{kugo}. 
In one version 
\be
m_{\nu}^D \approx m_u = m_t \left(\matrix{  0  & 0 & x \cr
                                            0 & x & 0 \cr
                                            x & 0 & 1 \cr} \right), ~~M 
\sim  M_0 \left(\matrix{ M_1 & M_2 & 0  \cr
                           M_2 & M_3 & 0  \cr
                           0   & 0   & M' \cr} 
\right), 
\ee
where $x \equiv m_c/m_t$, $M_{i} \sim M \sim 10^{10}$ GeV and 
$M' \sim 10^{16}$ GeV. 
Notice that mixing between the first and the second generations 
is small. It is interesting that the  Georgi- Yarlskog ansatz for charged
leptons and quarks leads to  
\be 
\sin \theta_{e \mu} \approx \frac{1}{3\sqrt{2}} \sin \theta_c \approx 0.05 
\ee
well in the range of small mixing 
MSW solutions of the solar neutrino problem.

Thus the large lepton mixing does not imply necessarily 
strong interfamily connection. As the consequence of the see-saw
mechanism, the large mixing 
can follow from small interfamily mixing (in $M$ and  $m^D$) 
but strong mass hierarchy in $M$.

\subsection{Large mixing and fine tuning?}

Large (maximal) mixing ``likes"  degeneracy. Indeed, 
the  symmetric $2 \times 2$ mass matrix with 
the diagonal elements $a$ and  $c$,  
and   the  off-diagonal elements $b$, leads to 
\be 
\tan \theta = \frac{2b}{a - c}, ~~~~~ 
\frac{m_1 - m_2}{m_1 + m_2} = 
\frac{2b}{a + c}~.   
\ee 
The second  equality holds for the large mixing angle 
which implies $(a - c)/b \ll 1$. 
If  $a/b \ll 1$ and $c/b \ll 1$, then one gets 
pseudo Dirac neutrino system with $m_1 \approx - m_2$ 
and $|m_1| -  |m_2| \approx (a + c)/2b$. 

In the scenario of fig.~1    
which explains both the solar and atmospheric neutrino data 
one encounters the following problem: 
$\nu_2 - \nu_3$ subsystem  should have large mixing and strong  
mass hierarchy: $m_2/m_3 \sim (5 - 6) \cdot 10^{-2}$. 
This means that all the elements of the mass matrix should be 
almost equal each other: $a \approx b \approx c$. In particular, 
$b \approx a~(1 - 2m_1/m_2)$.

In the context of the see-saw mechanism the large mixing and 
strong mass hierarhy can be reconciled if  

(i) only one right handed neutrino participates in the see-saw 
mechanism (or gives dominant contribution) and 

(ii) this RH neutrino couples equally with both LH components
\cite{josh,zurab}.

The mass matrix for this 
subsystem  is 
\be
\left(\matrix{ 0 & 0 & m_{1} \cr
               0 & 0 &  m_2 \cr
               m_{1} & m_2  & M \cr} \right) ~. 
\label{ber1}
\ee
It leads to one massless state and 
maximal mixing of the light components 
provided $ m_{1} \approx  m_2$.  
Small corrections to the above 
structure result in strong mass hierarchy of the eigenstates. 

The dominance 
of  only one RH neutrino contribution to the see-saw can be 
achieved in two different ways: 

(i) one of the RH neutrinos is much lighter than 
two others. This is equivalent to the see-saw enhancement
due to the strong mass hierarchy.  

(ii) The Yukawa coupling of one  RH neutrino with left components 
is much larger than the couplings of others RH neutrinos \cite{altar}. 
This leads to dominance of the corresponding 
two elements in the Dirac mass matrix \cite{altar}. 
In the latter case the mixing is enhanced by $m_{\nu}^D$.

\subsection{Large mixing from $m_l$} 

There is some hint that mixing between the second 
and third generations 
of leptons can be different from the quark mixing as well as  
mixing of the first and second generations. 
Indeed, 
\be
\frac{m_{\mu}}{m_{\tau}} = 3 \frac{m_s}{m_b} = 10 \frac{m_c}{m_t}. 
\label{hie}
\ee
The lepton hierarchy is weaker. 
Weak mass hierarhy can testify 
for larger mixing. So,  the enhancement of mixing 
can be  associated with the charged leptons.

The non-symmetric Dirac matrix has two off -diagonal  parameters 
which can control independently mass hierarchy and mixing, so that 
the problem of  fine 
tuning discussed in the  sect. 3.4 does not appear. 
Consider the following mass matrices in the basis 
$f_L^c M_F f_L$ (left components are  to the right):  
\be
m_l =  m_{\tau} \left(\matrix{  0  & \epsilon  \cr
                             \rho & 1         \cr} \right),~~~
m_d =  m_{b} \left(\matrix{  0  & \rho  \cr
                       \epsilon' & 1     \cr} \right),
\label{gularge}
\ee 
where $\rho \sim 1 \gg  \epsilon, \epsilon'$. 
Mass hierarchy is determined by the smallest off-diagonal
element.  
Since only the left component rotations contribute to the 
mixing,  the matrices $m_l$ and  $m_d$ lead to two different 
mixings of the quarks and leptons: 
The matrix $m_l$ is diagonalized by 
large angle rotation of the left components, 
whereas $m_d$ is diagonalized by small angle rotations of the left
components. 

Such a situation can be easily realized in the $SU_5$ 
Grand Unification, where the {\it left} components of 
the charged leptons, $l_L$,  are
unified  with the {\it right} components of the down quarks, $d_R$ in the 
{\bf 5}-plet, whereas right components $l_R$ are unified with $d_L$  
in the {\bf 10} - plets.     
Therefore one expects the same large rotations 
of the $l_L$ (which determine the mixing)  and $d_R$ (which is irrelevant
for mixing). The mass terms 
$$
h_{23} {\bf \bar{5}}_2 {\bf 10}_3 {\bf 5}_H + 
h_{32} {\bf \bar{5}}_3 {\bf 10}_2 {\bf 5}_H   
+ h_{33} {\bf \bar{5}}_3 {\bf 10}_3 {\bf 5}_H
$$
and $h_{23} \sim h_{33} \gg h_{32}$ reproduce  matrices (\ref{gularge}). 
(Here subscripts indicate the generation number.)  
\begin{figure}[htb]
\hbox to \hsize{\hfil\epsfxsize=6cm\epsfbox{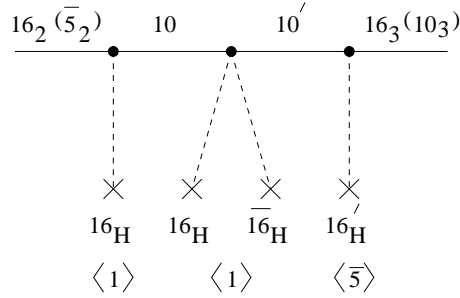}\hfil}
\caption{The diagram which generate large lepton mixing
}
\label{fdiag}
\end{figure}

In the $SO_{10}$ model,  ${\bf \bar{5}}$ and ${\bf 10}$-plets 
enter the same 
16-plet,  and consequently,  $h_{23} = h_{32}$. Therefore 
realization of the above possibility \cite{babu} implies the 
$SO_{10}$ breaking by the effective Yukawa couplings and asymmetry in 
the interactions of ${\bf 16}_2$ and ${\bf 16}_3$. 
For this new heavy  supermultiplets were introduced \cite{babu}: 
${\bf 10}$ and ${\bf 10'}$ of  matter fields   and  
${\bf 16}_H$, ${\bf 16}_H'$ ${\bf \bar{16}}_H'$ 
of Higgses. The  couplings 
$$
{\bf 16}_2 ~ {\bf 10}~ {\bf 16}_H  + 
{\bf 16}_3 ~{\bf 10'}~{\bf 16}_H'  
+ \frac{1}{M_P} 
{\bf 10}~{\bf 10'}~{\bf 16}_H~{\bf \bar{16}}_H' 
$$ 
generate the mass
terms via the diagram (fig. \ref{fdiag}). The Higgs doublet is in 
${\bf 16}_H'$. In 
the effective Yukawa coupling the $SO_{10}$ is broken 
by  VEV of ${\bf 16}_H$. 
The asymmetry appears since ${\bf 16}_2$ couples with ${\bf 16}_H$ 
which gets the VEV in the direction of singlet of $SU_5$, 
whereas  ${\bf 16}_3$ couples with ${\bf 16}_H'$ whose $SU_2$ doublet 
component gets a VEV (at the electroweak scale).

\subsection{Large mixing from $m^D_{\nu}$}

In the GU theories such as  $SU_5$ the Dirac mass terms 
for upper quarks, charged leptons and neutrinos have different gauge
structure: 
\be 
m_u \sim {\bf 10} \cdot {\bf 10} , ~~ 
m_d,~  m_l \sim {\bf \bar{5}} \cdot {\bf 10}, 
~~m_{\nu} \sim {\bf \bar{5}} \cdot{\bf 1}~.
\label{coupl}
\ee
Suppose the smallness of the mixing and the mass hierarchy
characterized by  small parameter $\epsilon$ is  associated 
with $10$-plet~\cite{BB}. Then according to (\ref{coupl}) one would expect 

\noindent
(1) hierarchy (smallness of mixing) of the order $\epsilon^2$ 
for the upper quarks,\\ 
(2) smallness $\epsilon$  for the down quarks and charged leptons 
(and indeed this is observed in  experiment!)\\ 
(3) no hierarchy of masses and mixings for neutrinos! 
Thus one expects large lepton mixing. 

This scenario can be realized, if the states in 10-plet 
are mixtures of the light and 
superheavy ($M \sim \Lambda_{GU}$) fermions \cite{BB}: 
\be 
u^0 \approx \epsilon u + U, ~~  d^0 \approx \epsilon d + D, ~~~
e^0 \approx \epsilon e + E. 
\ee  
So, the hierarchy of the masses and mixing follows from mixing 
in the 10-plet.\\ 

Thus, large lepton mixing is  
consistent with the  quark-lepton symmetry and 
the Grand Unification. The  difference between the 
quark and lepton 
mixing can be related to breaking of the left-right symmetry 
and breaking of the $GU$-symmetry itself.

\section{Family symmetry and large mixing}

The  observed mass and mixing hierarhy can be  
a consequence of the $U(1)_F$ family symmetry~\cite{FN}. 
Let us summarize  main points of the approach in~\cite{ramond}. 

1. Fermions carry certain $U(1)_F$ charges. 
(Prescription of charges which leads  
to realistic mass matrices implies that $U(1)_F$-symmetry 
is anomalous.)
For lepton doublets and RH neutrinos we denote the charges as 
\be
L_i:  (q_1, q_2, q_3), ~~~ N^c_i = (n_1, n_2, n_3)~~~ 
(i = 1, 2, 3)~.
\ee

2. $U(1)_F$ is spontaneously broken by 
non-zero VEV of the field $\theta$ with the charge $q_{\theta} = -1$. 
(In  version \cite{ross} two fields 
$\theta$ and $\bar{\theta}$ with opposite charges are introduced,  
see below).   

3. The Yukawa couplings appear as the effective operators 
after $U(1)_F$-symmetry breaking. For neutrinos we have:  
\be 
h_{ij} L_{i}H_u N_j^c 
\left(\frac{\langle \theta \rangle}{M}\right)^{q_i + n_j} + 
M_N \xi_{ij} N_i^c N_j^c
\left(\frac{\langle \theta \rangle}{M}\right)^{n_i + n_j}~,  
\label{yuk1}
\ee
where $h_{ij},  \xi_{ij} = O(1)$. 
$M$ is the scale at which the operators (\ref{yuk1}) are generated. 
In the renormalizable theory $M$ is the scale of masses of 
new heavy bosons or fermions which are integrated out in (\ref{yuk1}). 
It is suggested that $\langle \theta \rangle$ is smaller than 
$M$: 
\be 
\lambda  \equiv   \frac{\langle \theta \rangle}{M}  \sim
\sin \theta_c~,  
\ee 
where $\theta_c$ is the Cabibbo angle. It is this 
parameter $\lambda$ that determines the mass and mixing hierarchy. 

The couplings (\ref{yuk1}) generate  the mass matrices for neutrinos: 
\be 
\begin{array}{l}
m_{\nu}^D = diag(\lambda^{q_1}, \lambda^{q_2}, \lambda^{q_3})
~~\hat {h} ~~diag(\lambda^{n_1}, \lambda^{n_2}, \lambda^{n_3}) 
\langle H_u \rangle,\\
M_{\nu} = diag(\lambda^{n_1}, \lambda^{n_2}, \lambda^{n_3})
~~ \hat{\xi}~~
diag(\lambda^{n_1}, \lambda^{n_2}, \lambda^{n_3}) M_N~.
\end{array}
\ee
The see-saw formula  gives for light neutrinos: 
\be
m_{\nu} = \frac{\langle H_u \rangle ^2}{M} 
diag(\lambda^{q_1}, \lambda^{q_2}, \lambda^{q_3}) 
~\hat {h}~ \hat{\xi}^{-1}~ \hat {h}^{T} 
diag(\lambda^{q_1}, \lambda^{q_2}, \lambda^{q_3}).
\ee
Notice that $m_{\nu}$ does not depend on charges of the RH neutrinos.  
(In this approach the structures of the $m_{\nu}^D$ and $M$ 
are correlated.) 

The charges were assigned using the  phenomenological 
input, in particular,    
\be
\frac{m_s}{m_b} \approx  \lambda^2, ~~~  \frac{m_{\mu}}{m_{\tau}} \approx
\lambda^2, ~~~ V_{cb} \approx \lambda^2. 
\label{phenom}
\ee 
(In fact,  the above relations are satisfied 
with accuracy up to the factor 3.) This allows to reconstruct the 
mass matrix of down quarks,  using also the equality of charges 
$q_{ij} +  q_{ji} =  q_{ii} + q_{jj}$:  
\be
m_d \propto  
\left(
\matrix{  \lambda^4  & \lambda^3 & \lambda^3  \cr
          \lambda^3  & \lambda^2 & \lambda^2  \cr
          \lambda  &  1        &   1        \cr} \right)~.
\label{matrR}
\ee 
The key result which eventually leads to the possibility of 
large lepton mixing is that the elements of the second and the third 
columns are of the same order. This means that 
that $d^c_2$ and $d^c_3$ have the same 
charges and the  matrix (\ref{matrR}) is diagonalized by the large angle
rotation 
of these right components. 

The connection between the $U(1)_F$ charges of the quarks and leptons 
has been established in the following way.  
The charges of quarks can be written 
as $q(d^c_i) = B(2, -1 , -1)$, where $B$ is the baryon number. 
This expression can be generalized to include the leptons 
in spirit of $SU_5$ as $q(f_i) = (B - L)(2, -1 , -1)$ 
which gives for leptons $q(L_i) = - (2, -1, -1)$ \cite{ramond}. 
As in the quark sector,  the charges of the second and the third
generations of leptons are equal:  
$q_2 = q_3~$. 
The assignment of charges  leads to  the mass matrix of neutrinos: 
\be 
m_{\nu} \propto  \frac{\langle H_u \rangle ^2}{M}
\left(
\matrix{  \lambda^6  & \lambda^3 & \lambda^3  \cr
          \lambda^3  & a       &   b          \cr
          \lambda^3  & b       &   c          \cr} \right), 
\ee
where $a, b, c \sim O(1)$. 

This matrix is diagonalized by large $\nu_2 - \nu_3$ rotation. 
Therefore it 
admits  large $\mu - \tau$ mixing. At the same time,   
the symmetry does not give large mixing automatically. 
Indeed,  in this approach the LH components of the leptons 
of the second  and third generations also have the
same charges, and therefore,  the 
matrix elements of the  same order. As the consequence,  the 
charged lepton mass matrix is also diagonalized by 
large 2 - 3 rotation. Resulting lepton mixing 
is determined by mismatch of the  two large rotations  
of neutrinos and charge leptons and 
it may not be large.  Thus, explanation of the large mixing 
is reduced to a  theory  of  prefactors in front 
of powers of $\lambda$. 

As it was discussed in sect. 3.4,  simultaneous explanation of the solar
and the atmospheric neutrino 
problems implies $a \approx b \approx c$.   
So,  precise mass and mixing pattern depends on  
values of prefactors which are not determined in this approach. 

The equality of the charges  can  
follow  from  ``non-parallel" charge
assignment~\cite{yanagida} 
according to which the same charges have the 
following matter multiplets of the $SU(5)$:  
${\bf 10}_1$, $({\bf 10}_2, {\bf \bar{5}}_1)$ and $({\bf 10}_3, 
{\bf \bar{5}}_2,  {\bf \bar{5}}_3)$.\\

In a realization of the horizontal symmetry~\cite{ross} suggested
earlier,   
the $U(1)_F$ symmetry  
is  broken by VEV's of two fields  
$\theta$ and $\bar{\theta}$ with opposite charges. 
There are two small parameters which determine the 
mass hierarchy: 
$\epsilon \equiv \langle \theta \rangle/ M_2$
and 
$\bar{\epsilon} \equiv \langle \bar{\theta} \rangle/ M_2$.  
Down quarks and charged leptons have different 
charge prescriptions 
which is motivated by the weak hierarchy 
(\ref{hie}). 
As a consequence,  the mass matrix 
for the charge leptons has  the following 
hierarchical structure: 
\be
m_l \propto 
\left(
\matrix{  
0                 & \bar{\epsilon}^3    & 0                      \cr
\bar{\epsilon}^3  & \bar{\epsilon}      &  \sqrt{\bar{\epsilon}} \cr
0   &  \sqrt{\bar{\epsilon}}            &   1               \cr}\right)~,   
\ee
whereas  in the mass matrix for down quarks the 23-element 
is  $\sim \bar{\epsilon}$. The mixing between the second and third
generations equals  
\be
V_{23} = \sqrt{\bar{\epsilon}} + e^{i\delta} \epsilon~, 
\ee
where the  last term follows from diagonalization of the neutrino mass
matrix. 
The main contribution to mixing comes from the charge leptons.

\section{Democratic approach} 

One can generate hierarchies of masses and mixing 
starting by  the mass matrices 
of the  ``democratic"  form~\cite{dem}:  
\be
\hat{D} = 
\left(
\matrix{ 1  & 1 & 1  \cr
         1  & 1 & 1  \cr
         1  & 1 & 1  \cr} \right)~. 
\ee
The matrix has  the permutation $S_{3L} \times S_{3R}$ 
symmetry. The exact symmetry leads to zero mixing and only 
one nonzero mass eigenstate. This is the first approximation to the 
situation in the quark sector. Weak violation of the symmetry 
leads to the mass hierarchy and small mixing. 

Within this framework it is possible to explain 
large lepton mixing and avoid the fine tuning we have discussed 
in sect. 3.4.  
The main observation is that the Majorana mass matrix  has the symmetry
$S_{3L}$ which admits more general the form  
\be
m_{\nu} = a \hat{I} + b \hat{D}~, 
\ee
where $\hat{I}$ is the unit matrix. If for some reason 
({\it e.g.} related to the zero electric charge of neutrino) 
$b = 0$, then $m_{\nu} = a \hat{I}$. At the same time  
$m_l \propto \hat{D}$ which gives the following mixing matrix:  
\be 
V_{MNS} = 
\left(
\matrix{ 
\frac{1}{\sqrt{2}}  & - \frac{1}{\sqrt{2}} & 0                   \cr
\frac{1}{\sqrt{6}}  & \frac{1}{\sqrt{6}} & - \frac{2}{\sqrt{6}}  \cr
\frac{1}{\sqrt{3}}  & \frac{1}{\sqrt{3}} &  \frac{1}{\sqrt{3}}   \cr}
\right)~. 
\label{MNSFX}
\ee
In the limit of exact $S_{3L} \times S_{3R}$ two leptons are massless 
and only tau lepton gets the mass; 
all three neutrinos are degenerate. 
The weak violation of $S_{3L} \times S_{3R}$ gives   
small masses of muon  and electron  and   splitting of 
neutrino masses. 
Different forms of violation of the symmetry lead to   different 
phenomenological consequences.  
In \cite{tanim} the symmetry was broken by diagonal matrices 
$$
\delta m_{\nu} = diag(-\epsilon_{\nu}, \epsilon_{\nu}, \delta_{\nu}),
~~~~~ \delta m_l = diag(-\epsilon_l, \epsilon_l, \delta_l)~, 
$$ 
where $\epsilon$ and $\delta$ are fixed by masses of neutrinos and
charged leptons.  
They lead  to only weak 
modification of the mixing matrix. Instead of zero in (\ref{MNSFX}) 
one gets $- (2 \sqrt{{m_e}/ m_{\mu}})/ \sqrt{6}$. 

Such a scheme  can   
solve  the solar neutrino problem via ``Just-so"  oscillations  
$\nu_e \rightarrow \nu_{\mu}, \nu_{\tau}$ with 
$\sin^2 2\theta = 1$.  
The  atmospheric neutrino data 
are explained  by  $\nu_{\mu} \leftrightarrow \nu_{\tau}$ 
oscillations with 
$\sin^2 2\theta = 8/9$. 
Small admixture of $\nu_e$ in $\nu_3$ ($\sin^2 2\theta \approx 
16/6 m_e/m_{\mu} = 1.5\cdot 10^{-2}$) can be relevant for atmospheric
neutrinos 
and also can induce strong adiabatic transitions 
$\nu_e \rightarrow \nu_{\mu}, \nu_{\tau}$
in supernovae.  

In contrast, the symmetry violating matrix~\cite{FTY}
\be
\delta m_{\nu} =
m_0 \left(
\matrix{
0  & \epsilon_{\nu} & 0  \cr
\epsilon_{\nu}  & 0 & 0  \cr
0  & 0 & \delta_{\nu}  \cr}
\right) 
\label{FTY}
\ee
will lead to small $\nu_e - \nu_\tau$ mixing and 
$\Delta m_{12}^2 = 4 \epsilon_{\nu}m_0$,  
$\Delta m_{13}^2 = 2 \delta_{\nu}m_0$.

Notice that in this approach the origin of the large lepton mixing is the
Majorana character of neutrinos which implies $S_{3L}$ symmetry 
and the assumption that $m_{\nu} \propto \hat{I}$. 

In this approach the smallness of the neutrino mass is not 
explained, and the see-saw mechanism does not work. 
Indeed, the Dirac mass matrix of neutrinos  
$m_{\nu}^D \propto \hat D$ leads to $m_{\nu} \propto \hat D$ 
for any non-singular mass matrix of the RH components.  

\subsection{Universal strength of Yukawas }

According to ansatz~\cite{branco}  
the mass matrices have the form 
\be
m_{\nu} =
\left(
\matrix{
e^{i \alpha_1} & 1 & 1  \cr
1  & e^{i \alpha_2} & 1  \cr
1  & 1 & e^{i \alpha_3}   \cr}
\right)~. 
\label{branco}
\ee
Certain choice of phases leads to desired  mass hierarchies and
mixing. For instance, $\alpha_1 = \alpha_2$  for neutrinos  
and $\alpha_1 = - \alpha_2$ for charged leptons result in  
small mixing MSW solution of the solar neutrino problem and 
$\nu_{\mu} \leftrightarrow \nu_{\tau}$ oscillation solution 
of the atmospheric neutrino problem. \\

\section{Beyond Three Neutrinos}
\label{NNS}

There are two motivations for the introduction of
sterile neutrinos:
(i) to reconcile different neutrino anomalies
including  the LSND result;
(ii) to explain  existence of the large mixing in the
leptonic sector  (in contrast with quark sector).
Large mixing  implied by  the atmospheric neutrino
data can be  the mixing of  $\nu_{\mu}$  with sterile neutrino. All   
flavor mixings can be small.

There is another, indirect connection related to the fact that
large (maximal) mixing prefers degeneracy of mass. 
If the atmospheric neutrino problem is solved by oscillations of
$\nu_{\mu}$ and $\nu_{\tau}$ which strongly mix 
in degenerate states, then the only  way to solve the solar neutrino
problem without additional degeneracy is to introduce a sterile
neutrino which mixes with $\nu_e$.

\subsection{Schemes with  $\nu_{\mu} -  \nu_s$ oscillations of atmospheric
neutrinos}

Two possibilities have been discussed.

\begin{figure}[htb]
\hbox to \hsize{\hfil\epsfxsize=8cm\epsfbox{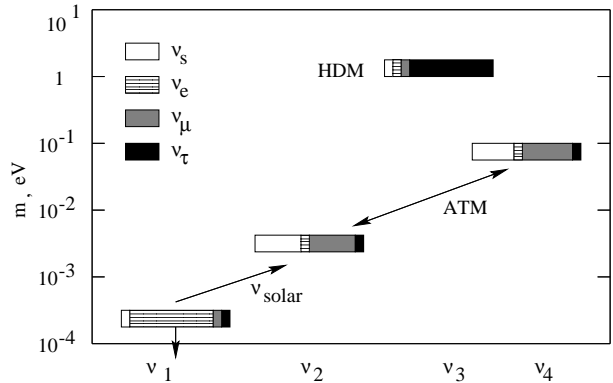}\hfil}
\caption{~~Neutrino masses and mixing in the
intermediate mass scale scenario. Here white parts of boxes
correspond to the sterile state.
}
\label{fint}
\end{figure}

I. {\it Intermediate mass scale scenario} is characterized by neutrino
mass hierarchy, small mixing,
and the Majorana masses of the  right  neutrinos
(in the context of the see-saw) at the intermediate mass scale:
$10^{10} - 10^{13}$ GeV. In addition, the light singlet fermion
can be  introduced  to solve the atmospheric neutrino problem
\cite{LS} (fig.~\ref{fint}).
The  neutrino masses equal
\be
m_4 = (0.3 - 3) \cdot 10^{-1} {\rm eV},~~~
m_2 \sim 3 \times 10^{-3}{\rm eV},~~~
m_3 \sim 1 {\rm eV}, ~~~
m_1 \ll  m_2.
\ee
$\nu_s$ and $\nu_{\mu}$ mix strongly  in the $\nu_2$ and $\nu_4$
eigenstates,
so that  $\nu_{\mu} \leftrightarrow  \nu_s$ oscillations
solve the  atmospheric neutrino problem;
$\nu_e \rightarrow \nu_{\mu}, \nu_{s}$ resonance conversion
explains the solar neutrino data, and $\nu_3$ can
supply significant amount of the HDM.\\

\begin{figure}[htb]
\hbox to \hsize{\hfil\epsfxsize=8cm\epsfbox{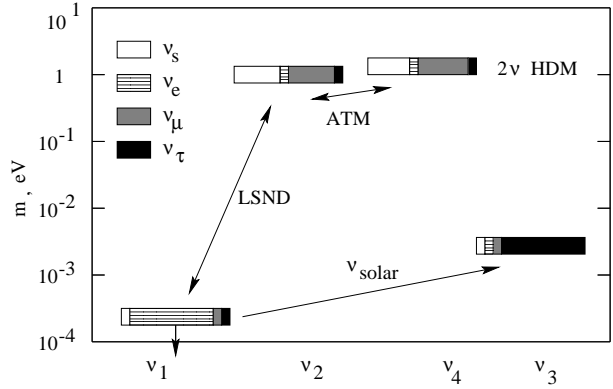}\hfil}
\caption{~~The neutrino masses and mixing
in the Grand Unification scenario.
}
\label{fgu}
\end{figure}

II. {\it Grand Unification Scenario}. 
The see-saw mechanism  based on the
Grand Unification   leads to the mass of the heaviest
neutrino ($\approx \nu_{\tau}$) in
the range $(2 - 3)\cdot 10^{-3}$ eV, and hence,  to a solution of the
solar 
neutrino problem via the $\nu_e \rightarrow \nu_{\tau}$ MSW conversion.
An  existence of the  light singlet fermion, $\nu_s$,  which
mixes predominantly with muon neutrino through the mixing mass
$m_{\mu s} \sim O(1)$ eV
allows one \cite{JS} (i) to solve the atmospheric neutrino problem
via the $\nu_{\mu} \leftrightarrow \nu_s$ oscillations, (ii) to
explain the LSND result and (iii) to get two component
hot dark matter in the Universe (fig.~\ref{fgu}). Similar scheme
has been suggested previously in another context \cite{GU}.

\subsection{On $\nu_{\mu} - \nu_s$ solution of the atmospheric 
neutrino problem}

There are two important features of the $\nu_{\mu} - \nu_s$ 
oscillation solution which distinguish it from the  
$\nu_{\mu} - \nu_{\tau}$  one. 

(i) Matter effect is important for the $\nu_{\mu} - \nu_s$  
mode and it does not influence practically 
$\nu_{\mu} - \nu_{\tau}$ mode.  
The matter effect changes total numbers and  distributions of 
the upward-going muon events and   
the multi-GeV events.  

The matter potential for the  $\nu_{\mu} - \nu_s$ system 
is determined by the concentration of neutrons 
$V = G_F n_n/\sqrt{2}$. 
For $\Delta m^2 < 10^{-2}$ eV$^2$ and  energies 
$E > 25$ GeV 
the zenith angle ($\Theta$)
dependence of the $\nu_{\mu} - \nu_s$ 
survival probability 
has rather peculiar form with two dips~\cite{LS,LMS,LL}.  

The  wide dip with minimum  at the zenith angles $\cos\Theta \sim - 0.4$, 
it reaches 1 at $\cos\Theta \sim - 0.8$, 
is due to resonance enhancement of oscillations in the mantle 
of the Earth. 
The second narrow dip  at $\cos\Theta \sim - 0.8 - 0.95$  
is due to the parametric enhancement of oscillations 
for neutrinos which cross both the core of the earth and the mantle 
of the Earth. The enhancement 
occurs when the phase of oscillations in the mantle and in the 
core equals $\pi$. The integration 
over the neutrino energies leads to significant flattening of the 
zenith angle dependence of the upwardgoing muons
\cite{LMS}. 
Shallow minima still can be seen in the vertical bins and 
at $\cos\Theta \sim - 0.6  \div  - 0.4$. Such a dependence 
differs from the one for the $\nu_{\mu} - \nu_{\tau}$ 
oscillations (where the matter effect is absent). It resembles 
the dependence in the MACRO experiment, where however,  the 
suppression in the vertical bins is stronger. 
Clearly, more data is needed to identify the solution. \\

(ii) The rate of the neutral current events should  be 
sustantially suppressed  
in the $\nu_{\mu} - \nu_s$ case, and the rate is not 
changed in the $\nu_{\mu} - \nu_{\tau}$ case. The best way to 
study  the NC effects is to detect the so called 
``$\pi^0$" - events \cite{suzuki,VS,pakw}: 
isolated $\pi^0$ which can be identified by the 
two gamma decay. 
Main contribution to ``$\pi^0$" - events comes 
from reaction 
\be 
\nu N \rightarrow \nu N \pi^0. 
\ee
At the SK, ``$\pi^0$" - events can also be  generated by the charged
current reactions {\it e.g. } 
$\nu_{\mu} N \rightarrow \mu  N' \pi^0$,  when $\mu$ is below the 
Cherenkov threshold (similar is for electron neutrinos). 
The $\nu_{\mu} -\nu_{\tau}$ oscillations only weakly 
suppress the number of ``$\pi^0$", 
whereas $\nu_{\mu} -\nu_s$ can suppress  the 
``$\pi^0$" rate by  30  - 40 \%. 

Preliminary SK data for 535 days give 210 ``$\pi^0$" events.  
This number exceeds the expected 192.5 events. 
To avoid  normalization uncertainties one can consider the 
ratio of numbers of the ``$\pi^0$" - events and the $e$-like events:  
$\pi^0/e$. The experiment gives \cite{SKat} 
\be
\frac{(\pi^0/e)_{data}}{(\pi^0/e)_{MC}} = 0.93 \pm 0.07 (stat) \pm 0.19
(syst.)
\label{dratio}
\ee 
which is consistent with both channels of oscillations. 
The double ratio is smaller than 1  
in spite of the  the excess of the $\pi^0$ due to even larger 
excess of the $e$-like  events. 
Large systematic error in (\ref{dratio}) is related to uncertainties in
the cross-sections. Notice that multi-pion production reactions
give significant contribution to the 
$\pi^0$ events (due to Cherenkov radiation threshold one or even more
pions are not detected). In \cite{VS} the total uncertainty 
was estimated as being at the level 30\%. 
The uncertainty will be diminished by direct measurements 
of the cross-section in the ``forward" detector of the long baseline 
experiment K2K \cite{suzuki}. 

Another uncertainty  is related to  
background, {\it e.g.},  from  interactions of neutrons 
in the detector. For the $\pi^0$ events this background 
is much more significant than for the  e-like events 
since only in 17\% of cases $\pi^0$ will induce the e-like event. 

It is also possible to study the zenith angle 
dependence of the $\pi^0$-events which is free of the 
uncertainties in the cross-section \cite{pakw}. 

Another suggestion is to study the up-down asymmetry  of the 
inclusive multiring events \cite{HM}.

Of course, the detection of the tau leptons produced 
by the $\nu_{\tau}$ would be direct way to 
identify the solution. However, the number of the expected events is
rather
small \cite{HM1}, and it is difficult  to reconstruct them. 

\subsection{Scheme with two degenerate neutrinos}
   
Maximal  mixing prefers  strong mass  degeneracy. Therefore
the  atmospheric neutrino result 
can be considered as an indication that
$\nu_{\mu}$ and $\nu_{\tau}$  are strongly mixed in the
two almost  degenerate neutrino states:
$\Delta m \ll   m_2 \approx m_3 \approx m_0$.
If $m_0 \sim 1$ eV,  these neutrinos can compose the
2$\nu$ HDM component in the Universe. In this case
the splitting should be $\Delta m \approx (2 - 5)\times 10^{-3}$ eV. The
first neutrino
composed, mainly,  of   $\nu_e$
can be much lighter: $m_1 \ll m_0$,  so that no observable signal in
the double beta decay is expected.

\begin{figure}[htb]
\hbox to \hsize{\hfil\epsfxsize=8cm\epsfbox{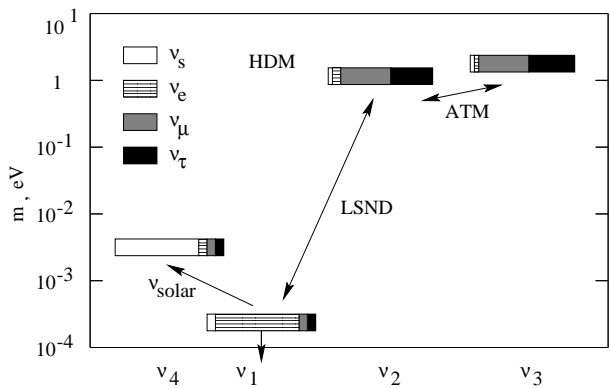}\hfil}
\caption{~~The pattern of the neutrino mass and mixing in the
scheme with two degenerate neutrinos and one sterile
component.
}
\label{fdeg}
\end{figure}

To explain the solar neutrino deficit one can introduce
sterile neutrino which mixes with $\nu_e$.
Then solar neutrinos undergo the $\nu_e \rightarrow \nu_s$
resonance conversion. 
This solution is characterized by weaker day-night effect but 
stronger distortion of the energy spectrum as compared with 
$\nu_e \rightarrow \nu_{\mu}$ conversion.   
The resulting scheme (fig.~\ref{fdeg}) can also explain the LSND 
result,  if the admixture of the $\nu_e$ in the heavy state
is large enough $U_{e3} \sim 2\times 10^{-2}$ (see~\cite{four}).

\section{On the Origin of the Sterile  Neutrinos} 

\subsection{Sterile neutrino or light singlet fermion?} 

From phenomenological point 
of view there is no difference between the sterile neutrino and 
light singlet fermion.  
The difference can be related to the origin of this state. 
We can keep the notion of the sterile neutrinos for the states which have 
generation structure. That is, one may expect that there are three 
sterile neutrinos. Such a possibility is realized in the  GU theories with
extended symmetry like $E_6$ \cite{MaE6}  where each fermionic generation  
(in {\bf 27}-plet) contains neutral fermion being a  singlet of $SO(10)$.  
Another realization is  the models with  
mirror symmetry~\cite{mirr}, where each neutrino 
has its mirror counterpartner. However,  even in these
cases it could be that 
not  all three additional neutrinos are light, and therefore 
number of light additional states is smaller than three.

Light singlet fermions have no generation structure. 
They  follow from 
some other sector of theory. 
Their number is not related to the
number of fermionic generations. 
It can be only one 
such a light singlet.  
For instance, axino or susy partner of the majoron 
can play the role of the singlet~\cite{CJS}.\\  

Common questions  are 

Why the singlet is so light ? 

How does it mix with usual neutrinos? 

A number of new possibilities has been considered recently.  

\subsection{Light singlet and supersymmetry} 

Some hint to the origin of the singlet may follow from 
the following numerology.  
In the supersymmetric theories there are additional mass 
scales which can determine  properties of the 
singlet fermion: a scale of the supersymmetry breaking
(the gravitino mass $m_{3/2}$), the $\mu$ parameter, 
the string scale and Planck scale $M_P$. 
The singlet may ``know" about these scales. 

Using  mass parameters $m_{3/2}$ (or $\mu$) and  $M_P$  one can compose
the following  mixing mass : 
\be
m_{\nu S} = \frac{m_{3/2}\langle H_2 \rangle }{M_P}, 
\label{smix}
\ee 
where $\langle H_2 \rangle$ is the VEV of the Standard model Higgs, 
and it is obviously needed to mix the doublet neutrino with 
singlet. The mass of the singlet can be constructed  as 
\be
m_{S} = \frac{m_{3/2}^2}{M_P}. 
\label{smass}
\ee 
It is interesting that for the supergravity value 
$m_{3/2} \sim$ 1 TeV the masses $m_{S}$ and $m_{\nu S}$ 
lead to oscillation parameters required for a  
solution of the solar neutrino problem via the 
MSW $\nu_e \rightarrow S$ resonance conversion \cite{BenSmi}. 
In particular, the mixing angle is determined simply by the 
ratio of the electroweak scale and the gravitino mass 
\be
\theta \sim \langle H_2 \rangle / m_{3/2}.  
\label{theta}
\ee
In the above example the mass parameters being  proportional to $m_{3/2}$ 
 appear when SUSY is broken. 

Alternatively, one can use 
supersymmetric $\mu$- parameter instead of $m_{3/2}$ 
\cite{BenSmi,Karim}. 
In this case $m_{S}$ and $m_{\nu S}$ exist even in the
supersymmetric limit. This opens a possibility 
to realize the scenario in models with  gauge mediated SUSY breaking,  
where  $m_{3/2}$ is small \cite{Karim}.

In \cite{BenSmi} it was suggested that $S$ originates from 
the Hidden sector of  theory. 
One possibility is that $S$ is the modulino -- 
supersymmetric partner of the moduli field~\cite{BenSmi}. 
In this case the R-parity should be broken.\\

If $S$ is a  non-moduli field, its properties can be determined by
additional $U(1)$ gauge factor which is broken at TeV scale \cite{BenSmi}. 

This idea has been elaborated recently in~\cite{Lang}. Mass terms 
of the sterile neutrino are generated by the non-renormalizable 
effective interactions with another Standard Model singlet $S'$ charged 
under additional $U(1)$. The following effective couplings 
have been introduced in the 
superpotential: 
\be 
L S H_2 \left(\frac{S'}{M_s}\right)^{p_D} + 
S^{T}S S' \left(\frac{S'}{M_s}\right)^{p_M} + ...~~. 
\label{super}
\ee
The parameters ${p_D}$ and ${p_M}$ are determined by $U(1)$ 
charges of $S$ and $S'$; $M_s$ is the string scale. 
Additional singlet $S'$ has the following  potential 
\be 
\frac{1}{2} m_{soft}^2 S'^2 + A \left(\frac{S'^{2 + k}}{M_s^k}\right)^2~,
\ee
where the last term 
comes from the non-renormalizable interactions in the
superpotential and $m_{soft} \sim$ TeV. Minimization gives 
\be
\langle S' \rangle \sim (m_{soft} M_s )^{1/(k+1)}~.
\ee
When $S'$ acquires the VEV, the interactions (\ref{super}) 
generate the Dirac and the Majorana masses: 
\be
m_{D} = \langle H_2 \rangle 
\left(\frac{m_{soft}}{M_s}\right)^{p_D/(k+1)}
\label{smix1}
\ee
\be
m_{M} = m_{soft} 
\left(\frac{m_{soft}}{M_s}\right)^{(p_M - k)/(k+1)}~.
\label{smass1}
\ee
If $p_d = p_M - k$, 
then $m_{D} / m_{M} \sim \langle H_2 \rangle / m_{soft}$ 
similarly to (\ref{theta}). 

From practical point of view the most interesting example is when 
$k = 1$, $p_D = 2$ and $p_M = 3$. It gives 
$m_{D} = \langle H_2 \rangle m_{soft}/{M_s}$ and 
$m_{M} = m_{soft}^2/{M_s}$ as in (\ref{smix},\ref{smass}). 
For $m_{soft} \sim 1$ TeV they  lead to solution of the 
solar neutrino problem. To solve the atmospheric 
neutrino problem via $\nu_e \leftrightarrow \nu_s$ oscillations 
one need rather exotic values of parameters: 
$k = 6$, $p_D = 5$ and $p_M = 12$ which imply 
high dimension nonrenormalizable terms 
in the superpotential. 

\subsection{Composite fermions as sterile neutrinos}

New possible origin of the light singlet fermions  
has been suggested in~\cite{Grossman}. It is assumed that there is 
a new sector of the theory which includes preons and  
gauge interactions with strong  dynamics at some scale 
$\Lambda$. 
This dynamics leads to confinement of preons. Moreover,  
it is assumed 
that the dynamics leaves  unbroken a chiral symmetry 
and therefore generates massless composite baryons. 
These composite states are identified with sterile neutrinos. 

An example of $SU(n + 4)$ ($n \geq 1$) gauge theory has been elaborated. 
Massless baryons have the structure  
\be 
B_{ij} = \frac{1}{\Lambda^3} \psi_i A \psi_j~,  
\ee
where $A$ is antisymmetric tensor and 
$\psi_i$ are antifundamentals $(i = 1 ... n)$. Such a theory is shown 
to produce $n(n + 1)/2$ massless baryons.

The interactions of preons with the SM particles is realized via 
high dimensional operators suppressed by power of the mass scale 
$M$. Moreover, $M \gg \Lambda$ and it is this inequality 
leads to smallness of the 
mixing mass. In particular,  the following terms  
were introduced: 
\be
\lambda^{i j \alpha} \frac{1}{M^3}\psi_i A \psi_j L_{\alpha} H^{\dagger}
= \left(\frac{\Lambda}{M}\right)^3 \lambda^{i j \alpha} \psi_i A \psi_j
L_{\alpha} H^{\dagger}~,
\label{mix2}
\ee
where $L_{\alpha}$ and $H$ are  the SM leptonic doublet and Higgs. 
The mass of $B_{ij}$ can be also generated by the 
non-renormalizable operators which break chiral symmetry:  
\be   
h^{i j k l} \frac{1}{M^5} \psi_i A \psi_j \psi_k A \psi_l   
= h^{i j k l} M \left(\frac{\Lambda}{M}\right)^6 B_{ij} B_{kl}. 
\label{mass2}
\ee 
From (\ref{mix2},\ref{mass2}) we get: 
$m_{\nu S} \sim \lambda \langle H \rangle (\Lambda / M)^3$, 
$m_{S} \sim M (\Lambda / M)^6$. For $M \sim 10^{18}$ GeV and 
$\Lambda \sim 10^{13}$ GeV the neutrino masses are in the 
range of small mixing MSW solution of the solar neutrino problem.


\section{Summary}

1). Large lepton mixing  
can be well  consistent with our standard notions: 
hierarchy of masses weak interfamily connection, the see-saw mechanism.  
It  can  be just artifact of the 
see-saw mechanism (see-saw enhancement). 

\noindent
2). The large lepton mixing can however be an indication of physics beyond
our standard notion. Still it 
is consistent with unification of 
quarks and leptons. 

Large  mixing in $\nu_{\mu} - \nu_{\tau}$ channel 
can be naturally reconciled with small mixing in other channels. 

\noindent
3). An extreme point of view is that 
large lepton mixing is  the mixing of muon neutrino 
with new state --  sterile neutrino.  
This possibility can be checked by studies of the 
$\pi^0$ events and the  zenith angle dependence  of the
upward going muons. 

\noindent
4). The introduction of sterile neutrino can be motivated 
by explanation of large lepton mixing. 
This allows  one to keep all flavor mixings to be small, 
and to rescue natural scenarios of neutrino mass and mixing. 

Large $\nu_{\mu} - \nu_{\tau}$ mixing can be naturally 
associated with mass degeneracy. In  such a scheme, the solar 
neutrino problem is  solved via the 
$\nu_{e}- \nu_s$ conversion. 
The latter can be checked by studies of 
correlation of the spectrum distortion and the day-night effect at SK 
and in future by studies of the neutral current 
interactions  in Sudbury Neutrino Observatory.

\noindent
5). The number of the phenomenological schemes of neutrino masses and
mixing is rather restricted now. 
Clear signatures exist 
for each scenario. The key steps in reconstructing the 
neutrino mass spectrum are: 

(i) identification of  the 
solutions of the atmospheric neutrino problem:  
$\nu_{\mu} - \nu_{\tau}$ or  
$\nu_{\mu} - \nu_{s}$;  

(ii) clarification of the role of the subdominant mode 
$\nu_{\mu} - \nu_{e}$ in the oscillations of atmospheric neutrinos;   

(iii) identification of the solution of the 
solar neutrino problem: Just- so, MSW or may be something else?

(iv) clarification of a  role of sterile neutrinos in  conversion of
the solar neutrinos.

\section{Note added}

Materials included in this review have been published or reported 
before the Symposium (middle of June, 1998). 
Since that time large number of new 
papers in the field has been published. 
Below we  give some relevant references. 

1. Phenomenology of various  scenarios of neutrino mass and mixing have
been studied in \cite{BGG,Raffelt,Bar,Ahl}. 

2. A number of publications \cite{Lola,Ellis,Viss,Alt} 
is devoted to 
properties of the mass matrices (textures, symmetries) 
which lead to explanation of the solar 
and atmospheric neutrino problems. 
Scenario with bi-maximal mixing is of special interest  
(for the phenomenological aspects see {\it e.g.}~\cite{CG}). 
There are several attempts to construct the model 
which naturally leads to bi-maximal mixing. 
Gauge model with four generations and 
certain discrete symmetry has been suggested in 
~\cite{MoNU}. Another version~\cite{MoNU1} is based on the 
left-right symmetric gauge model with additional 
$S_3\times Z_4 \times Z_3 \times Z_2$ symmetry. 
See also~\cite{KK}. The bi-maximal mixing 
has  been also  considered  in the 
MSSM with single RH neutrino~\cite{DK} as well as in   $SO_{10}$ GUT
\cite{NY}.

3. Consequences of the   $U(1)$ flavor symmetry 
for  the neutrino mass matrix and lepton mixing 
were further elaborated in~\cite{Josh,JR,FGN,CCH}. 
It is argued that large lepton mixing can be 
a natural consequence of the $U(2)$ flavor 
symmetry~\cite{HW}. 

4. The ``democratic approach" has been  
summarized in~\cite{MTanimoto}. As is shown in~\cite{Fukugita}   
the democratic mass matrices can be ``embedded"  
in the Grand Unified  $SU(5)$ model.  
The scenario with three degenerate neutrinos has been 
discussed in~\cite{Ma1}.  

5. Aspects 
Generation of large lepton mixing has been considered in 
minimal version of the see-saw  
mechanism~\cite{MaRS} in   
GUT~\cite{Berezh}, in  models with 
radiative mechanism~\cite{EM}, and in  
composite model~\cite{Haba}.    

New possibilities of 
the description of neutrino data in  models 
with gauge mediated SUSY breaking have been 
studied in \cite{AnjanJ}.

6. A number of papers is devoted to generation of neutrino mass in
supersymmetric models via  the R-parity violating 
interactions~\cite{JF,Broo,MRV,JChun,RBR}. 

7. Phenomenology of schemes with more than 3 light neutrinos 
and properties of the corresponding mass matrices were  
discussed in~\cite{BPWW,Mohanty,Schwetz,Lipmanov}. 

There are new attempts to construct a (3 + 1)- model  (three active 
neutrinos and one sterile  neutrino)  
based on the radiative mechanism~\cite{Oka} 
as well as  singular see-saw~\cite{CLJS}. 

8. New mechanism for generation of the light sterile neutrino  
in the supersymmetric model with gauge mediated 
SUSY breaking~\cite{DvaliNir} has been suggested.  
The see-saw model of sterile neutrino was considered in~\cite{Brah}.

9. Completely new possibilities to explain smallness 
of neutrino mass and the 
lepton number violation as well as appearance of the light singlet states 
are based on existence of large extra dimensions~\cite{A-HD,ADD,DDG},  
see also~\cite{Ibanez}.


\section*{Acknowledgments}
I would like to thank Cy Hoffman and Peter Herczeg for hospitality.


\section*{References}
\begin{thebibliography}{99}

\bibitem{SKat}Y. Fukuda et al., Phys. Rev. Lett., {\bf 81} (1998) 1562.  

\bibitem{SKsun} Y. Suzuki, Talk given at Int. Conf. Neutrino-98. 
18th Int. Conf. on Neutrino Physics and Astrophysics
(NEUTRINO 98), Takayama, Japan, 4-9 Jun 1998.   

\bibitem{LSND}C. Athanassopoulos, Phys. Rev. Lett. {\bf 81} (1998) 1774; 
W. Louis, these Proceedings. 

\bibitem{KAR}G. Drexlin, these Proceedings, K. Eitel et al.,
18th Int. Conf. on Neutrino Physics and Astrophysics
(NEUTRINO 98), Takayama, Japan, 4-9 Jun 1998,  
hep-ex/9809007. 


\bibitem{HDM} E. Gawiser and J. Silk, Science {\bf 280} (1998) 1405;
J. Primack and M. A. K. Gross, astro-ph/9810204. 

\bibitem{CHOOZ} M. Apollonio et al., Phys. Lett. B{\bf 420} (1998) 397 

\bibitem{MY} H. Minakata and O. Yasuda, Nucl. Phys., {\bf B 523} (1998)
597. 

\bibitem{mkkee}R. P. Thun and S. McKee, hep-ph/9806534. 

\bibitem{bimax} See e.g. F. Vissani,  hep-ph/9708483; V. Barger, S.   
Pakvasa, T. Weiler and K. Whisnant, hep-ph/9806387, A. Baltz,
A. S. Goldhaber and M. Goldhaber, hep-ph/9806540, H. Georgi
and S. L. Glashow,  hep-ph/9808293.

\bibitem{three} P. F. Harrison, D. H. Perkins, W. G. Scott,
Phys.  Lett. {\bf  B396}   (1997) 186. 

\bibitem{Peccei} See e.g. R.D. Peccei, K. Wang, Phys. Rev. {\bf D53} 
(1996)  2712, M. Matsuda, M. Tanimoto, Phys.Rev. {\bf D58} (1998) 093002.  

\bibitem{FY} M. Fukugita, M. Tanimoto, T. Yanagida 
Prog. Theor. Phys. {\bf 89} 263 (1993).  

\bibitem{barshay}S. Barshay, P. Heiliger, Astropart. Phys. {\bf 6} 
323  (1997), J. Pati, Talk given at the    
18th Int. Conf. on Neutrino Physics and Astrophysics
(NEUTRINO 98), Takayama, Japan, 4-9 Jun 1998. 

\bibitem{tao} A. Yu.  Smirnov and Z. Tao, (in preparation).  

\bibitem{zee} A. Zee, Phys. Lett, {\bf 93 B} (1980) 389. 

\bibitem{TS} A. Yu. Smirnov, M. Tanimoto, Phys. Rev. {\bf D55} 1665
(1997) 1665, N. Gaur, A. Ghosal, E. Ma, 
P. Roy, Phys. Rev. {\bf D58} 071301 (1998). 

\bibitem{seesaw} M. Gell-Mann, P. Ramond, R. Slansky, in Supergravity,
ed. by F. van Nieuwenhuizen and Freedman (Amsterdam,
North Holland, 1979) 315; T. Yanagida, in Proc. of the
workshop on the unified Theory and Baryon Number in the Universe,
eds. O. Sawada and A. Sugamoto (KEK, Tsukuba, 1979) 95.
R. Mohapatra and G. Senjanovi\'c,  Phys. Rev. Lett.
44 (1980) 912.


\bibitem{MNS} Z. Maki, M. Nakagawa, S. Sakata, Prog. Theor. Phys. {\bf 28} 
(1962) 247.

\bibitem{AS}A. Yu. Smirnov,  Nucl. Phys. {\bf B466} 25 (1996). 


\bibitem{sees} A. Yu. Smirnov, Phys. Rev. {\bf D48} 3264 (1993);  
M. Tanimoto, Phys. Lett. {\bf B345} (1995) 477. 


\bibitem{kugo} M. Bando, T. Kugo, K. Yoshioka, Phys. Rev. Lett. {\bf 80} 
(1998) 3004. 

\bibitem{josh}A. S. Joshipura, Phys. Rev. {\bf D51} 1321 (1995). 

\bibitem{zurab}Z. G. Berezhiani and A. Rossi, Phys. Lett. 
{\bf B367} (1996); B. C. Allanach hep-ph/9806294.

\bibitem{altar}G. Altarelli and F. Feruglio, hep-ph/9807353. 

\bibitem{babu}C. H. Albright, K. S. Babu and S. M. Barr, hep-ph/9805266. 

\bibitem{BB}K. S. Babu and S. M. Barr, Phys. Lett, {\bf B381} (1996) 202.

\bibitem{FN}C. Froggatt, H. B. Nielsen, Phys. Lett. {\bf B147} (1979) 227. 
M. Leurer, Y. Nir,  N. Seiberg, Nucl. Phys. {\bf B398} 319 (1993). 

\bibitem{ramond}J. K. Elwood, N. Irges and P. Ramond hep-ph/9807228. 

\bibitem{ross}G. K. Leontaris,  S. Lola, G.G. Ross, Nucl. Phys. {\bf B454}
(1995) 25; 
D. Dreiner et al., Nucl. Phys., {\bf B436} (1995) 461. 

\bibitem{yanagida}J. Sato and T. Yanagida, hep-ph/9809307. 

\bibitem{dem}H. Fritzsch and Z. Xing, 
Phys. Lett. {\bf B440} (1998) 313;  hep-ph/9808272 and references
therein. 

\bibitem{tanim}M. Fukugita, M. Tanimoto, T. Yanagida
Phys. Rev. {\bf D 57} (1998) 4429. 

\bibitem{FTY}
M. Fukugita, M. Tanimoto, T. Yanagida, hep-ph/9809554. 

\bibitem{branco}
G. C. Branco, M. N. Rebelo, J. I. Silva-Marcos, 
Phys. Lett. {B428} 136 (1998).

\bibitem{LS} Q. Y. Liu, A. Yu. Smirnov,  Nucl. Phys. {\bf B 524} (1998)
505.

\bibitem{JS} A. S. Joshipura, A. Yu. Smirnov, hep-ph/9806376.

\bibitem{GU}
J. T. Peltoniemi, D. Tomasini and J. W. F. Valle, Phys. Lett. {\bf B298}
(1993); E. J. Chun, C. W. Kim and U. W. Lee, hep-ph/9802209.


\bibitem{LMS} Q. Y. Liu, S.P. Mikheyev, A.Yu. Smirnov, 
Phys. Lett. {\bf B440}  319  (1998), hep-ph/9803415.  

\bibitem{LL} P. Lipari and  M. Lusignoli, Phys. Rev. {\bf D58}   073005
(1998). 


\bibitem{suzuki}Y. Suzuki, Proc. of the Int. Conf. Neutrino-96, 
Ed. K. Enquist, K Huitu, J. Maalampi,  p. 237.   


\bibitem{VS} F. Vissani, A. Yu. Smirnov,
Phys. Lett. {\bf B432} 376 (1998). 

\bibitem{pakw}J.  G. Learned, S. Pakvasa, J. L. Stone
Phys. Lett. {\bf B435} 131 (1998).




\bibitem{HM}L. J. Hall, H. Murayama, Phys. Lett. {\bf B436} (1998) 323.  

\bibitem{HM1}L. J. Hall, H. Murayama, hep-ph/9810468.


\bibitem{four}
J.T. Peltoniemi and J.W.F. Valle,
Nucl. Phys. B {\bf 406},(1993) 409;
D.O. Caldwell and R.N. Mohapatra, Phys. Rev.{\bf D48}, (1993) 3259;
Z. Berezhiani and R.N. Mohapatra, Phys. Rev. \textbf{D52}, 6607 (1995);
E. Ma and P. Roy,  Phys. Rev. {\bf B52}, R4780 (1995);
E.J. Chun {\it et al.}, Phys. Lett. \textbf{B357}, (1995) 608;
R. Foot and R.R. Volkas,
Phys. Rev.  \textbf{D52}, (1995) 6595;
J.J. Gomez-Cadenas and M.C. Gonzalez-Garcia,
Z. Phys. C \textbf{71}, (1996) 443;
E. Ma, Mod. Phys. Lett. A \textbf{11}, (1996) 1893;
S. Goswami, Phys. Rev. {\bf D55}, (1997) 2931;
V. Barger, T.J. Weiler and K. Whisnant, hep-ph/9712495.


\bibitem{MaE6} E. Ma and P. Roy, Phys. Rev. {\bf D52}  (1995) 4780; 
E. Ma, Phys. Lett.  {\bf B380} (1996) 286; Mod. Phys. Lett. {\bf A11} 
(1996) 1898. 

\bibitem{mirr} Z. G. Berezhiani and R. N. Mohapatra, Phys. Rev. 
{\bf D52} (1995) 6607. 

\bibitem{CJS} E. J. Chun, A. S. Joshipura and A. Yu. Smirnov, 
Phys. Rev. {\bf D54} (1996) 4654. 


\bibitem{BenSmi} K. Benakli and A. Yu. Smirnov, 
Phys. Rev. Lett. {\bf 79} (1997)  4314, hep-ph/9703465.

\bibitem{Karim} K. Benakli, hep-ph/9801303. 

\bibitem{Lang} P. Langacker, J. Wang, 
Phys. Rev. {\bf D58} 115010, 1998, hep-ph/9804428.  

\bibitem{Grossman} N. Arkani-Hamed, Y. Grossman, hep-ph/9806223. 



\bibitem{BGG} 
S.M. Bilenkii, C. Giunti, W. Grimus,  
hep-ph/9809368, hep-ph/9807568, 9812360.

\bibitem{Raffelt} Georg G. Raffelt, hep-ph/9807484. 

\bibitem{Bar} V. Barger, T.J. Weiler, K. Whisnant, Phys. Lett. 
{\bf B440} (1998) 1, hep-ph/9807319.  

\bibitem{Ahl} D.V. Ahluwalia, Mod. Phys. Lett. {\bf A13} (1998)  2249, 
hep-ph/9807267.


\bibitem{Lola} S. Lola, J.D. Vergados, Prog. Part. Nucl. Phys. {\bf 40} 
(1998) 71. 

\bibitem{Ellis} J. Ellis, G.K. Leontaris, S. Lola, D.V. Nanopoulos, 
hep-ph/9808251.

\bibitem{Viss} F. Vissani, hep-ph/9810435. 

\bibitem{Alt} G. Altarelli,  F. Feruglio, hep-ph/9809596.

\bibitem{BHS} R. Barbieri, L. J. Hall, A. Strumia, hep-ph/9808333.  


\bibitem{CG} C. Giunti, hep-ph/9810272. 

\bibitem{MoNU} R. N. Mohapatra, S. Nussinov,  hep-ph/9808301.
  
\bibitem{MoNU1} R. N. Mohapatra, S. Nussinov,  hep-ph/9809415.  

\bibitem{KK} Sin Kyu Kang, C. S. Kim, hep-ph/9811379. 

\bibitem{DK} S. Davidson, S. F. King, hep-ph/9808296. 

\bibitem{NY} Y. Nomura, T. Yanagida, 
Phys. Rev. {\bf D59} (1999) 017303,  hep-ph/9807325.



\bibitem{Josh} A. S. Joshipura, hep-ph/9808261.

\bibitem{JR} A. S. Joshipura and S. D. Rindani, hep-ph/9811232. 

\bibitem{FGN} C. D. Froggatt, M. Gibson, H. B. Nielsen, 
hep-ph/9811265. 

\bibitem{CCH} K. Choi, E. J. Chun, K. Hwang, hep-ph/9811363. 


\bibitem{HW} L. J. Hall, N. Weiner, hep-ph/9811299. 



\bibitem{MTanimoto} M. Tanimoto, hep-ph/9807517.


\bibitem{Fukugita} M. Fukugita, M. Tanimoto, T. Yanagida, hep-ph/9809554.  


\bibitem{Ma1} E. Ma, hep-ph/9812344. 



\bibitem{MaRS} E. Ma, D.P. Roy, U. Sarkar, hep-ph/9810309. 

\bibitem{Berezh} Z. Berezhiani, A. Rossi, hep-ph/9811447. 
 
\bibitem{EM} Ernest Ma, hep-ph/9807386. 

\bibitem{Haba} N. Haba, hep-ph/9807552.

\bibitem{AnjanJ} A. S. Joshipura, S. K. Vempati, hep-ph/9808232. 

\bibitem{JF} J. Ferrandis, hep-ph/9810371.

\bibitem{Broo} G. Brooijmans, hep-ph/9808498.

\bibitem{MRV} B. Mukhopadhyaya, S. Roy, F. Vissani, hep-ph/9808265. 

\bibitem{JChun} E. J. Chun, S. K. Kang, C. W. Kim, U. W. Lee, 
hep-ph/9807327. 


\bibitem{RBR} S. Rakshit, G. Bhattacharyya, A. Raychaudhuri, 
hep-ph/9811500. 



\bibitem{BPWW}B. V. Barger, S. Pakvasa, T.J. Weiler, K. Whisnant, 
Phys. Rev. D58 (1998) 093016,  hep-ph/9806328.

\bibitem{Mohanty} S. Mohanty, D. P. Roy,  U. Sarkar, 
hep-ph/9808451. 

\bibitem{Schwetz} S. M. Bilenkii, C. Giunti, W. Grimus, T. Schwetz 
hep-ph/9807569. 

\bibitem{Lipmanov} E. M. Lipmanov, Phys. Lett. {\bf B439} (1998), 119, 
hep-ph/9806442.


\bibitem{Oka} Y. Okamoto, M. Yasue, hep-ph/9812403.

\bibitem{CLJS} Chun Liu, J. Song, hep-ph/9812381. 



\bibitem{DvaliNir} G. Dvali Y. Nir, 
J. High Energy Phys. 9810:014, 1998,  hep-ph/9810257. 

\bibitem{Brah} B. Brahmachari,  hep-ph/9812216. 



\bibitem{A-HD} N. Arkani-Hamed, S.  Dimopoulos, hep-ph/9811353.  

\bibitem{ADD} N. Arkani-Hamed, S. Dimopoulos, Gia Dvali, J. March-Russell
hep-ph/9811448.  

\bibitem{DDG}K. R. Dienes, E. Dudas, T. Gherghetta, hep-ph/9811428

\bibitem{Ibanez} L. E. Ibanez, C. Munoz, S. Rigolin, hep-ph/9812397.  


\end{thebibliography}
\end{document}